# Exploiting Ultralow Loss Multimode Waveguides for Broadband Frequency Combs

*Xingchen Ji,[1] Jae K. Jang,[2] Utsav D. Dave,[1] Mateus Corato-Zanarella,[1] Chaitanya Joshi,[2] Alexander L. Gaeta,[2] and Michal Lipson[1,*]*

[1]Department of Electrical Engineering, Columbia University, New York, NY, 10027, USA

[2]Department of Applied Physics and Applied Mathematics, Columbia University, New York, NY, 10027, USA

Corresponding Author: E-mail: ml3745@columbia.edu

**Abstract:**

Low propagation loss in high confinement waveguides is critical for chip-based nonlinear photonics applications. Sophisticated fabrication processes which yield sub-nm roughness are generally needed to reduce scattering points at the waveguide interfaces in order to achieve ultralow propagation loss. Here, we show ultralow propagation loss by shaping the mode using a highly multimode structure to reduce its overlap with the waveguide interfaces, thus relaxing the fabrication processing requirements. Microresonators with intrinsic quality factors (Q) of 31.8 ± 4.4 million are experimentally demonstrated. Although the microresonators support 10 transverse modes only the fundamental mode is excited and no higher order modes are observed when using nonlinear adiabatic bends. A record-low threshold pump power of 73 µW for parametric oscillation is measured and a broadband, almost octave spanning single-soliton frequency comb without any signatures of higher order modes in the spectrum spanning from 1097 nm to 2040 nm (126 THz) is generated in the multimode microresonator. This work provides a design method that could be applied to different material platforms to achieve and use ultrahigh-Q multimode microresonators.

Low propagation loss in high confinement waveguides which has enabled chip-based nonlinear photonics applications[1–10] is typically achieved by reducing scattering points at the waveguide interfaces, which requires complex fabrication processes. Up to now, the scattering points at the interfaces have been minimized solely through fabrication processes such as optimized etching[11–13], reflowing resist masks[14,15], and chemical mechanical planarization (CMP)[12,15–17]. However, it is challenging to develop an optimized fabrication process for different material platforms to achieve ultra-smooth interfaces, as complex adjustment of process parameters is required. Ultra-fine polishing pads and polishing slurries, gas optimization and tens of hours of high temperature resist reflow are generally needed to provide ultra-smooth interfaces[15–17].

Here we show low propagation loss in high confinement waveguides by shaping the mode using a highly multimode structure to reduce its overlap with the waveguide interfaces, thus relaxing the need to physically reduce the scattering points via fabrication processes. We show a design of a highly multimode microresonator that minimizes the interaction between the fundamental mode and excited higher order modes in order to prevent the increase of threshold for nonlinear processes[12,18]. Our design ensures propagation solely in the fundamental mode by preventing the excitation of higher order modes, in contrast to previous designs that ensure propagation solely in the fundamental mode by eliminating the excited higher order modes using loss[19]. The microresonator consists of a highly multimode waveguide with a bending radius that changes nonlinearly along the circumference from a large value (900 µm) in the coupling section to a small value (80 µm) in the bend (see Fig. 1(a)). The large value in the coupling section ensures efficient coupling and excitation of only the fundamental mode, while the small value allows fundamental mode propagation with a small effective bending radius. We engineer the microresonator by leveraging nonlinear functions for minimizing rapid changes in curvature, which have been well studied in mathematics and applied in photonics[20–23]. We design the bending radius to change nonlinearly in order to ensure adiabaticity while maintaining a small footprint. For comparison, a design with a bending radius that changes linearly - using the same slope as the one in the coupling section of the nonlinear case to ensure adiabaticity - would require a 10 times larger footprint. We choose the hyperbolic tangent (*tanh*) function for changing the bending radius as our nonlinear function to provide a high degree of adiabaticity with simple design implementation and a compact footprint. This *tanh* function ensures that the radius changes slowly in the coupling section, where higher order mode excitation is most sensitive. Note that in principle other nonlinear functions could also be used to change the radius along the circumference while ensuring adiabaticity.

We design a microresonator that supports 10 transverse modes while ensuring that approximately 95% of the input power circulating the resonator is in the fundamental mode when using nonlinear adiabatic bends. The microresonator shown in Fig. 1(a) has a free spectral range (FSR) of 174 GHz and is based on a waveguide with a cross-section of 730 nm x 2600 nm which supports 10 modes (Transverse electric (TE) modes, shown in Fig. 1(a) inset). The bus waveguide consists of the same cross-section and in order to ensure that only the fundamental mode is launched into the microresonator, it is terminated with an inverse taper. In Fig. 1(b), we show the full 3D finite-difference time-domain (FDTD) simulations (Lumerical FDTD) of the coupling section. In Fig. 1(c), we show the normalized power in each mode normalized by the total power coupled from the bus waveguide to the resonator. The simulations show that at 1550 nm in our resonator design the power in the higher order modes is almost 7 dB lower than that in the conventional resonator with a constant bending radius.

We show experimentally that intrinsic quality factors (Q) of 31.8 ± 4.4 million are achieved by propagating a single mode in the highly multimode microresonators. In order to test these devices, we use a lensed fiber to launch a laser source through a fiber polarization controller into the inverse nanotaper input of our chip and use a collimating lens to collect the transmitted light from the inverse nanotaper output of our chip. In Fig. 2(a) we show the measured transmission spectrum and the measured average intrinsic linewidth is 6.1 MHz ± 0.9 MHz, corresponding to an intrinsic Q of 31.8 ± 4.4 million and a propagation loss of less than 1 dB/m. The histogram of intrinsic linewidths in Fig. 2(b) shows that the low propagation loss is consistent across multiple FSRs and 4 different fabricated samples. In order to compare our designed resonator with a conventional resonator having a constant bending radius, we fabricate both types of structures on the same wafer. By choosing a constant bending radius of 131 µm, we ensure that both structures exhibit the same FSR. In Fig. 2(c), we show the measured normalized transmission spectra for the two structures. While for the microresonator with constant radius the transmission dips at multiple frequencies , indicating the presence of higher order mode excitation and propagation[24], for the microresonator with nonlinear adiabatic bends the transmission dips only at every FSR , indicating the lack of higher order modes. Note that in our experiment, the maximum intensity variation of the background is about 4%, which is therefore the sensitivity limit of higher order mode measurement.

We measure a record-low pump power threshold of 73 µW for parametric oscillation using the multimode microresonator and generate a broadband, almost octave spanning single soliton frequency comb without any fingerprint of higher order modes, spanning from 1097 nm to 2040 nm (126 THz). We measure the output power in the first generated four-wave-mixing sideband for different pump powers to determine the pump power threshold for parametric oscillation. The measured first sideband power for a pump wavelength around 1560 nm is shown in Fig. 3(a) and an example of the initial state of parametric oscillation with pump power of 73 µW is shown in Fig. 3(b). In order to measure the generated combs, we use two optical spectrum analyzers (OSAs) to acquire the spectrum due to the limited wavelength range of a single OSA. In Fig. 4(a), we show a broadband frequency comb spectrum generated with the microresonator before phase-locking and in Fig. 4(b), we show frequency comb single soliton spectrum generated with the microresonator close to octave spanning and phase-locked using the microheater thermal tuning method discussed in[25]. The fit to a single soliton state with a spectral $sech^2$ envelope is shown (in purple) in Fig. 4(b) and matches well with our experimental results. The comb spectra do not exhibit the typical defects that indicates the presence of higher order modes, such as missing lines or dips. Note that in Fig. 4(b) the spectrum exhibits dips (at 1350 nm and 1850 nm, for example) only at the frequencies corresponding to the narrow wavelength division multiplexing (WDM) used to filter the pump to increase the dynamic range of the OSA for the phase-locked comb.

Our demonstration of broadband frequency combs shows our ability to engineer the dispersion despite the large waveguide dimensions. The simulated group-velocity dispersion (GVD) parameter as a function of the microresonator width, with a fixed height of 730 nm for the fundamental TE mode and a pump wavelength of 1550 nm, is shown in Fig. 4(c). The dispersion in these wide structures is robust to fabrication variations and the shaded area indicates the anomalous GVD region needed for broadband soliton combs. For our experiment, we choose 2600 nm width (marked with the red star), because it minimizes the overlap between propagating mode and scattering points at the waveguide interfaces while providing small anomalous GVD favorable for broadband comb generation[26,27]. GVDs in these wide structures are also more robust to fabrication variations.

In conclusion, we show that a highly multimode microresonator can be used to achieve ultra-low propagation loss (< 1 dB/m) using as-deposited films, in contrast to previous demonstrations where ultra-low propagation loss (< 1 dB/m) is solely achieved by for example applying CMP to smoothen the deposited films[16,17]. Our method of using highly multimode structures to achieve ultra-low propagation loss eliminates the requirement for CMP and can be applied not only to $Si_3N_4$, but also to different material platforms[28–37], where developing optimized processes to achieve smooth interfaces is more challenging. In principle, these highly multimode structures could be further combined with sophisticated optimized fabrication processes to achieve extremely low propagation loss. Utilizing the nonlinear adiabatic design to efficiently excite and propagate only the fundamental mode increases flexibility in the waveguide design while ensuring ultra-low propagation loss as well as dispersion engineering. The generated unlocked, smooth and broadband comb with small line spacing could be useful for applications that do not require low phase noise, such as optical coherence tomography, while the generated phase-locked comb could benefit applications that rely on microresonators such as spectroscopy, metrology, astronomy, high-speed communications, and on-chip optical clocks.

**Acknowledgements**

This work was performed in part at the Cornell NanoScale Facility, a member of the National Nanotechnology Coordinated Infrastructure (NNCI), which is supported by the National Science Foundation (Grant NNCI-2025233). The authors acknowledge support from the Defense Advanced Research Projects Agency (HR0011-19-2-0014), the Air Force Office of Scientific Research (FA9550-15-1-0303, FA8650-19-C-1002), National Science Foundation (OMA-1936345). We would like to thank Dr. Janderson Rocha Rodrigues and Dr. Aseema Mohanty for helpful discussion. Dr. Xingchen Ji is also grateful to the China Scholarship Council for financial support.

**Reference:**

[1] T. J. Kippenberg, A. L. Gaeta, M. Lipson, and M. L. Gorodetsky, *Science* **361**, (2018).

[2] A. L. Gaeta, M. Lipson, and T. J. Kippenberg, *Nat. Photonics* **13**, 158–169 (2019).

[3] D. T. Spencer, T. Drake, T. C. Briles, J. Stone, L. C. Sinclair, C. Fredrick, Q. Li, D. Westly, B. R. Ilic, A. Bluestone, N. Volet, T. Komljenovic, L. Chang, S. H. Lee, D. Y. Oh, M.-G. Suh, K. Y. Yang, M. H. P. Pfeiffer, T. J. Kippenberg, E. Norberg, L. Theogarajan, K. Vahala, N. R. Newbury, K. Srinivasan, J. E. Bowers, S. A. Diddams, and S. B. Papp, *Nature* **557**, 81–85 (2018).

[4] A. Dutt, C. Joshi, X. Ji, J. Cardenas, Y. Okawachi, K. Luke, A. L. Gaeta, and M. Lipson, *Sci. Adv.* **4**, e1701858 (2018).

[5] E. Obrzud, M. Rainer, A. Harutyunyan, M. H. Anderson, J. Liu, M. Geiselmann, B. Chazelas, S. Kundermann, S. Lecomte, M. Cecconi, A. Ghedina, E. Molinari, F. Pepe, F. Wildi, F. Bouchy, T. J. Kippenberg, and T. Herr, *Nat. Photonics* **13**, 31–35 (2019).

[6] H. Bao, A. Cooper, M. Rowley, L. Di Lauro, J. S. Totero Gongora, S. T. Chu, B. E. Little, G.-L. Oppo, R. Morandotti, D. J. Moss, B. Wetzel, M. Peccianti, and A. Pasquazi, *Nat. Photonics* **13**, 384–389 (2019).

[7] X. Ji, X. Yao, A. Klenner, Y. Gan, A. L. Gaeta, C. P. Hendon, and M. Lipson, *Opt. Express* **27**, 19896–19905 (2019).

[8] B. Stern, X. Ji, Y. Okawachi, A. L. Gaeta, and M. Lipson, *Nature* **562**, 401–405 (2018).

[9] H. Lee, T. Chen, J. Li, K. Y. Yang, S. Jeon, O. Painter, and K. J. Vahala, *Nat. Photonics* **6**, 369–373 (2012).

[10] M.-G. Suh, Q.-F. Yang, K. Y. Yang, X. Yi, and K. J. Vahala, *Science* **354**, 600–603 (2016).

[11] M. Soltani, S. Yegnanarayanan, and A. Adibi, *Opt. Express* **15**, 4694–4704 (2007).

[12] H. E. Dirani, L. Youssef, C. Petit-Etienne, S. Kerdiles, P. Grosse, C. Monat, E. Pargon, and C. Sciancalepore, *Opt. Express* **27**, 30726–30740 (2019).

[13] A. Frigg, A. Boes, G. Ren, I. Abdo, D.-Y. Choi, S. Gees, and A. Mitchell, *Opt. Express* **27**, 37795 (2019).

[14] A. Gondarenko, J. S. Levy, and M. Lipson, *Opt. Express* **17**, 11366 (2009).

[15] M. H. P. Pfeiffer, J. Liu, A. S. Raja, T. Morais, B. Ghadiani, and T. J. Kippenberg, *Optica* **5**, 884 (2018).

[16] X. Ji, F. A. S. Barbosa, S. P. Roberts, A. Dutt, J. Cardenas, Y. Okawachi, A. Bryant, A. L. Gaeta, and M. Lipson, *Optica* **4**, 619 (2017).

[17] J. Liu, E. Lucas, A. S. Raja, J. He, J. Riemensberger, R. N. Wang, M. Karpov, H. Guo, R. Bouchand, and T. J. Kippenberg, *Nat. Photonics* 1–6 (2020).

[18] T. Herr, V. Brasch, J. D. Jost, I. Mirgorodskiy, G. Lihachev, M. L. Gorodetsky, and T. J. Kippenberg, *Phys. Rev. Lett.* **113**, 123901 (2014).

[19] A. Kordts, M. H. P. Pfeiffer, H. Guo, V. Brasch, and T. J. Kippenberg, *Opt. Lett.* **41**, 452–455 (2016).

[20] M. Cherchi, S. Ylinen, M. Harjanne, M. Kapulainen, and T. Aalto, *Opt. Express* **21**, 17814 (2013).

[21] T. Chen, H. Lee, J. Li, and K. J. Vahala, *Opt. Express* **20**, 22819 (2012).

[22] L. Yao, T. A. Birks, and J. C. Knight, *Opt. Express* **17**, 2962 (2009).

[23] L. Zhang, L. Jie, M. Zhang, Y. Wang, Y. Xie, Y. Shi, and D. Dai, *Photonics Res.* **8**, 684–689 (2020).

[24] Y. Liu, Y. Xuan, X. Xue, P.-H. Wang, S. Chen, A. J. Metcalf, J. Wang, D. E. Leaird, M. Qi, and A. M. Weiner, *Optica* **1**, 137 (2014).

[25] C. Joshi, J. K. Jang, K. Luke, X. Ji, S. A. Miller, A. Klenner, Y. Okawachi, M. Lipson, and A. L. Gaeta, *Opt. Lett.* **41**, 2565 (2016).

[26] S. Coen, H. G. Randle, T. Sylvestre, and M. Erkintalo, *Opt. Lett.* **38**, 37–39 (2013).

[27] Y. Okawachi, M. R. E. Lamont, K. Luke, D. O. Carvalho, M. Yu, M. Lipson, and A. L. Gaeta, *Opt. Lett.* **39**, 3535–3538 (2014).

[28] L. Razzari, D. Duchesne, M. Ferrera, R. Morandotti, S. Chu, B. E. Little, and D. J. Moss, *Nat. Photonics* **4**, 41–45 (2010).

[29] A. G. Griffith, R. K. W. Lau, J. Cardenas, Y. Okawachi, A. Mohanty, R. Fain, Y. H. D. Lee, M. Yu, C. T. Phare, C. B. Poitras, A. L. Gaeta, and M. Lipson, *Nat. Commun.* **6**, (2015).

[30] Y. He, Q.-F. Yang, J. Ling, R. Luo, H. Liang, M. Li, B. Shen, H. Wang, K. Vahala, and Q. Lin, *Optica* **6**, 1138–1144 (2019).

[31] G. Moille, L. Chang, W. Xie, A. Rao, X. Lu, M. Davanço, J. E. Bowers, and K. Srinivasan, *Laser Photonics Rev.* **14**, 2000022 (2020).

[32] Z. Gong, A. Bruch, M. Shen, X. Guo, H. Jung, L. Fan, X. Liu, L. Zhang, J. Wang, J. Li, J. Yan, and H. X. Tang, *Opt. Lett.* **43**, 4366–4369 (2018).

[33] D. J. Wilson, K. Schneider, S. Hönl, M. Anderson, Y. Baumgartner, L. Czornomaz, T. J. Kippenberg, and P. Seidler, *Nat. Photonics* **14**, 57–62 (2020).

[34] X. Shen, R. C. Beltran, V. M. Diep, S. Soltani, and A. M. Armani, *Sci. Adv.* **4**, eaao4507 (2018).

[35] D. T. H. Tan, K. J. A. Ooi, and D. K. T. Ng, *Photonics Res.* **6**, B50–B66 (2018).

[36] L. Chang, W. Xie, H. Shu, Q.-F. Yang, B. Shen, A. Boes, J. D. Peters, W. Jin, C. Xiang, S. Liu, G. Moille, S.-P. Yu, X. Wang, K. Srinivasan, S. B. Papp, K. Vahala, and J. E. Bowers, *Nat. Commun.* **11**, 1331 (2020).

[37] Z. Ye, A. Fülöp, Ó. B. Helgason, P. A. Andrekson, and V. Torres-Company, *Opt. Lett.* **44**, 3326–3329 (2019).

**Figures**

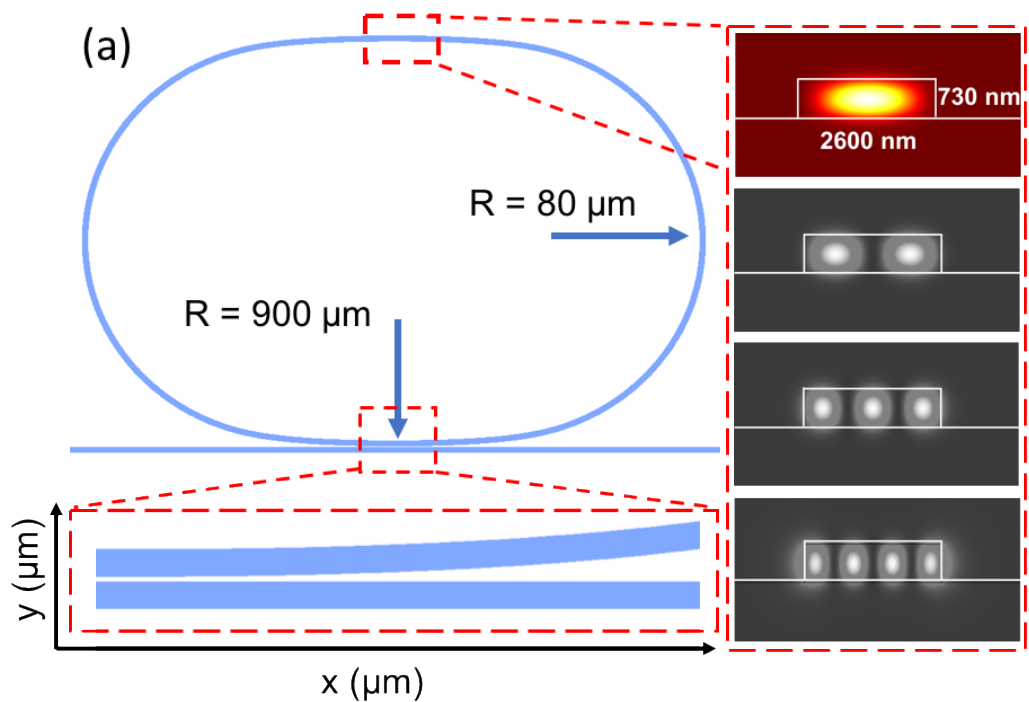

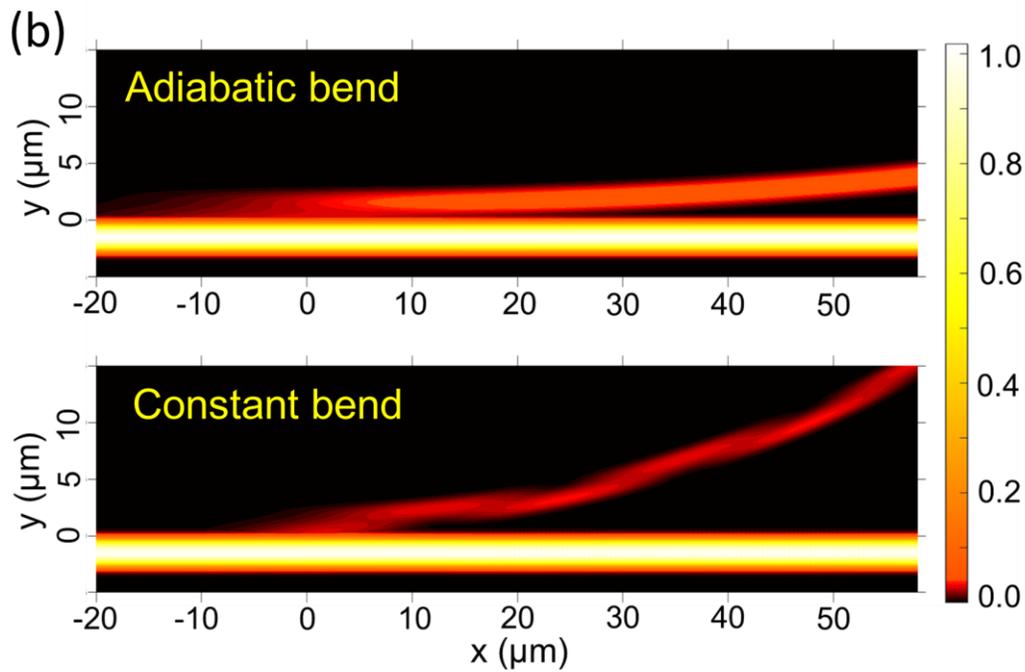

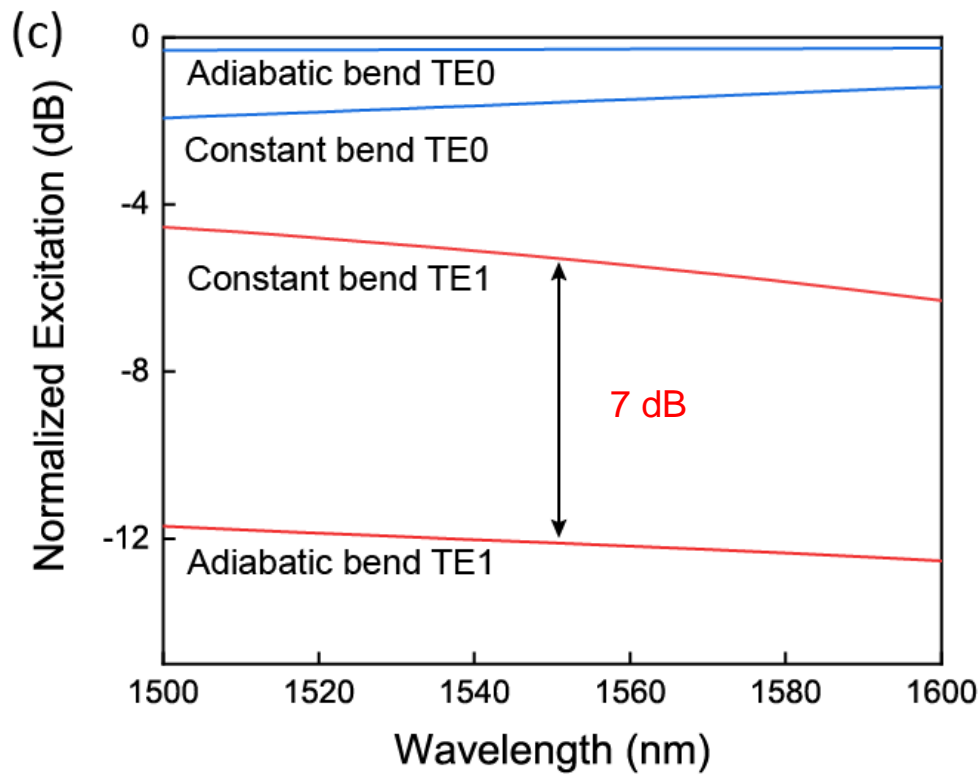

**Fig. 1**. (a) Schematic of our microresonators with adiabatic bends design. The bending radius is 900 µm in the coupling section and then gradually reduces to 80 µm in the sharpest bend. Inset shows the transverse electric (TE) modes supported by the waveguide and only the fundamental mode is excited in the adiabatic bends design. (b) FDTD simulations of the adiabatic bends design (top) and the conventional ring resonator with constant bending radius (bottom). Note that higher order modes are excited in the constant bending radius ring and not in our adiabatic bends design. (c) Simulations of normalized mode excitations for adiabatic bend and conventional bend with constant bending radius of 131 µm as a function of wavelength.

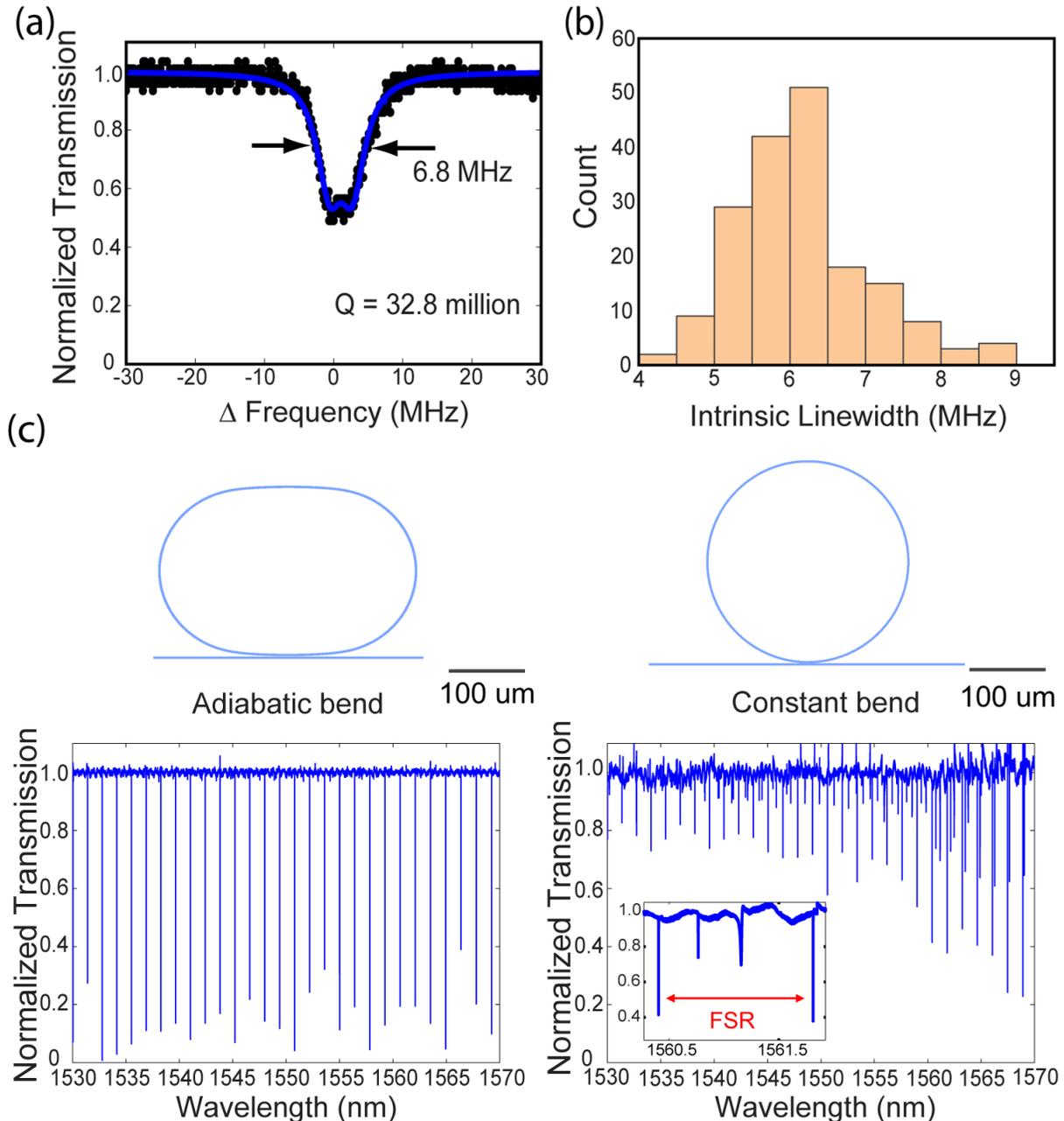

**Fig. 2**. (a) Normalized transmission spectrum of a resonance with an intrinsic Q of 32.8 million. (b) Histogram of intrinsic linewidths measured with several samples. The average intrinsic linewidth is 6.1 MHz ± 0.9 MHz. (c) Schematic of a microresonator with adiabatic bend design (top left). Measured normalized transmission spectrum with no higher order mode observed (bottom left). Schematic of microresonators with constant bending radius design (top right). Measured normalized transmission spectrum with clear signatures of higher order modes (bottom right). The inset shows the higher order modes resonance dips in between one FSR of the fundamental mode. Note that both microresonators have the same FSR and similar footprint.

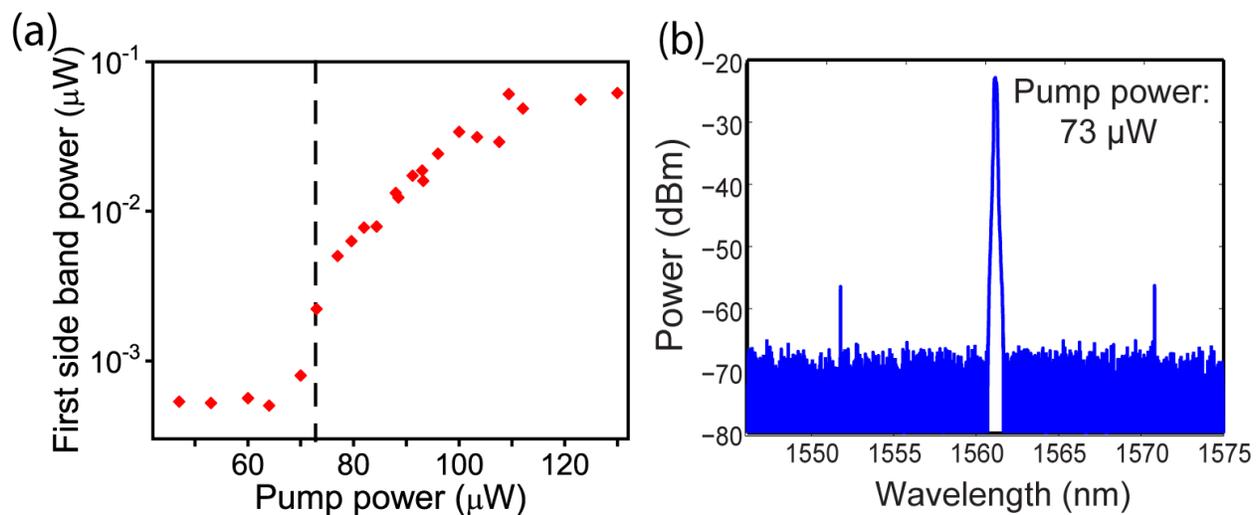

**Fig. 3**. (a) Output power in the first generated side band as a function of pump power. In this device, parametric oscillation occurs for pump power as low as 73 µW (indicated by the dashed vertical line) with a pump wavelength at 1561 nm. (b) Initial state of parametric oscillation measured with pump power of 73 µW.

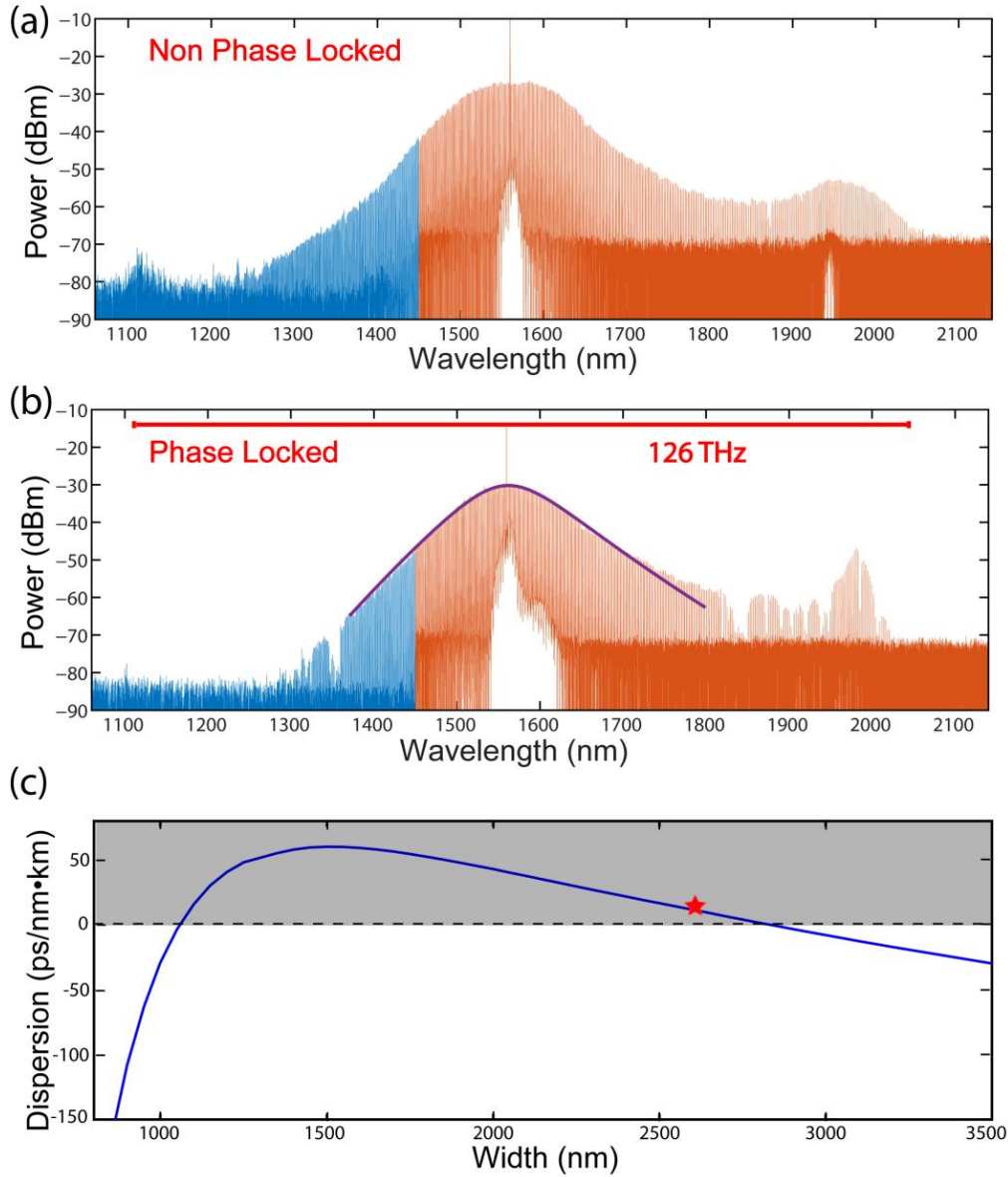

**Fig. 4**. (a) Broadband frequency comb before phase locking generated with our highly multimode microresonator. The comb is very smooth and does not have missing lines or dips, which indicates that no mode crossings happen and only the fundamental mode exists. (b) A phase-locked single soliton frequency comb close to octave spanning. The broadband spectrum spans from 1097 nm to 2040 nm (126 THz) with a free spectral range (FSR) of 174 GHz. Note that blue and red parts are the same single soliton state measured by two OSAs. The dips in the spectrum (at 1350 nm and 1850 nm for example) are due to the WDM filter, not due to mode crossings. The fit of single soliton state with a spectral $sech^2$ envelope is shown in purple and it matches very well with our experimental result. (c) Simulated group-velocity dispersion (GVD) parameter as a function of the microresonator width, with a fixed height of 730 nm for the fundamental TE mode. The shaded area indicates the anomalous GVD region needed for broadband spanning Kerr frequency combs. For our experiment, we choose 2600 nm width (marked with the red star), because it minimizes the overlap between propagating mode and scattering points at the waveguide interfaces while providing small anomalous GVD favorable for broadband comb generation.